\documentclass[conference]{IEEEtran}
\IEEEoverridecommandlockouts
\usepackage{cite}
\usepackage{amsmath,amssymb,amsfonts}
\usepackage{algorithmic}
\usepackage{graphicx}
\usepackage{textcomp}
\usepackage{xcolor}
\def\BibTeX{{\rm B\kern-.05em{\sc i\kern-.025em b}\kern-.08em
    T\kern-.1667em\lower.7ex\hbox{E}\kern-.125emX}}
\begin{document}

\title{Brain-like features of MemComputing machines\\
\thanks{This work was supported by the National Science Foundation under Grant ECCS-2229880.}
}

\IEEEoverridecommandlockouts
\IEEEpubid{\makebox[\columnwidth]{~
		\copyright~2023 IEEE \hfill} \hspace{\columnsep}\makebox[\columnwidth]{}}

\author{\IEEEauthorblockN{Massimiliano Di Ventra, {\it Fellow, IEEE}}
\IEEEauthorblockA{\textit{Department of Physics} \\
\textit{University of California, San Diego}\\
La Jolla, CA 92093, USA \\
diventra@physics.ucsd.edu}
}

\maketitle

\begin{abstract}
MemComputing is a new model of computation that exploits the non-equilibrium property---we call ``memory''---of any physical system to respond to external 
perturbations by keeping track of how it has reacted at previous times. Its digital, scalable version maps a finite string of symbols into a finite 
string of symbols. In this paper, I will discuss some analogies of the operation of MemComputing machines---in general, and digital in particular---with a few physical properties of the biological brain. These analogies could be a source of inspiration to improve on the design of these machines. In turn, they could 
suggest new directions of study in (computational) neuroscience. 
\end{abstract}
\vspace{.2cm}

\section{Introduction}
Although the complete understanding of how the biological brain works is still far from complete, some of its features have been experimentally determined. These include, for instance, the main components and operation of a single neuron, the establishment of short- and long-term memories~\cite{neuro}, the long-range correlations in the dynamics of clusters of neurons~\cite{Munoz}, etc. Neuro-inspired computation, when implemented in solid-state hardware~\cite{Meade} or emulated in artificial neural networks~\cite{Goodfellow}, aims at reproducing and/or exploiting some of these features for some tasks such as learning, association, pattern recognition, prediction, and so on. 

Recently, a new model of computation has been introduced, named {\it MemComputing}~\cite{diventra13a,UMM,Mybook}, which takes advantage of a physical property shared 
by all physical systems: {\it memory} (another name for ``time non-locality''). This {\it non-equilibrium} property means that when the state of a physical system is perturbed, the perturbation affects the system's state at a later time~\cite{Mybook,Membook}. Of course, in some cases the memory decays so fast that it is not experimentally detectable, or it is too small to be technologically useful. However, there are situations in which time non-locality is strong enough that it can be exploited for computing and that is the main idea behind MemComputing~\cite{Mybook}.

In this paper I will discuss some analogies MemComputing machines (in general, and digital in particular) share with the operation of the biological brain. Some 
of these analogies are {\it emergent phenomena} of the dynamics of these machines. Therefore, they could shed further light on the operation of the biological brain 
itself.

\section{Universal MemComputing machine}
Let us start from the general definition of the {\it universal} MemComputing machine (UMM)~\cite{UMM}.\footnote{The term ``universal'' means that this machine is 
	Turing-complete~\cite{UMM}.} A UMM is defined as the eight-tuple~\cite{UMM}
\begin{equation}
UMM=(M,\Delta,{\cal P},S,\Sigma,p_0,s_0,F)\,.\label{UMMdef}
\end{equation}
Here, $M$ indicates the set of possible states of a single memprocessor (the fundamental unit of a MemComputing machine). The set ${\cal P}$ contains the arrays of pointers, $p_\alpha$, that select the memprocessors called by the {\it transition function} $\delta_\alpha$. The set of indexes is indicated with $S$. $\Sigma$ is the set of initial states written by the input device on the computational memory. $p_0\in {\cal P}$ is the initial array of pointers, $s_0$ is the initial index $\alpha$, and $F\subseteq M$ is the set of final states.

The set of {\it transition functions}, $\Delta$, has elements 
\begin{equation}
	\delta_\alpha:M^{m_\alpha}\backslash F\times {\cal P}\rightarrow M^{m\rq_\alpha}\times {\cal P}^2\times S\,,\label{functUMM}
\end{equation}
with a number $m_\alpha<\infty$ of input memprocessors (read by the transition function $\delta_\alpha$), and $m\rq_\alpha<\infty$ output  memprocessors written by the same transition function. 

I note first that it was shown in~\cite{UMM} that the mathematical definition of the UMM encompasses also the description of artificial
neural networks (ANNs). In other words, ANNs can be viewed as a {\it special case} of MemComputing machines. However, Eqs.~(\ref{UMMdef}) 
and~(\ref{functUMM}) allow us to go beyond this realization and show additional similarities with the operation of the biological brain. These are as follows.

\subsection{Massively parallel architecture with combined information processing and storage}
Any transition function $\delta_\alpha$ of a UMM simultaneously acts on a set of memprocessors at once. This was named ``intrinsic
parallelism'' in~\cite{UMM} and it is fundamentally different from the ``standard parallelism'' of our modern computers (or even parallel Turing machines). Instead, this feature seems to belong to the biological brain, or at the very least, its representation as an artificial neural network~\cite{kohonen_book}. 

In addition, by construction, memprocessors and their network ({\it computational memory}) can process and store information simultaneously. Although it is still not fully clear how the brain performs these two tasks, compelling evidence points to the collection of neurons and synapses in the brain as the main agents able to concomitantly carry out these functions~\cite{neuro}. 

\subsection{Asynchronous computation}

Asynchronous computation means that all or a large chunk of processing units of a machine compute and exchange information simultaneously,
without the need to wait for a predetermined period
of time, such as the global clock period in our modern computers. Asynchronous computation is a main feature of the biological brain, and models of artificial neural networks. It also follows from the general definition of MemComputing machines; cf. Eqs.~(\ref{UMMdef}) and~(\ref{functUMM}). These machines do not require a global clock, and the different memprocessors compute and exchange information simultaneously.  

\subsection{Information overhead}

From the definition of a UMM as a collection of memprocessors it is clear that the {\it topology} of such a network needs to be specified and represents a fundamental aspect of these machines. This property has been called {\it information overhead}~\cite{UMM} and it is a type of ``data compression'' which is embedded in the machine at the outset of the computation, and does not vary during dynamics. Different types of information overheads (topologies) can be assumed ranging 
from polynomial to exponential~\cite{UMM}. It is indeed this property that allows a UMM with exponential information overhead to solve NP-complete problems with polynomial resources~\cite{UMM,Mybook}. (Note that this statement 
does {\it not} imply that NP=P, since MemComputing machines are {\it not} Turing machines.)

A similar concept has been discussed in the context of the brain~\cite{kohonen_book}. In fact, the brain appears to have a high level 
of specificity, in both the types of neurons and their network architecture (the connectome), even at a mesoscopic level. This indicates that the brain physical topology is not completely random. Rather, it shows some degree of specialization (information overhead) with 
important functional properties~\cite{neuro}. 

\subsection{Functional polymorphism}

The set $\Delta$ of transition functions (\ref{functUMM}) of a UMM may contain more than a single element. This means that a UMM can, in principle, compute {\it different functions} without modifying the topology of the machine network, by simply applying the appropriate input signals. This feature was named {\it functional polymorphism}~\cite{UMM}, and it is not available to our modern computers (or Turing machines). A practical realization of this concept was reported in~\cite{DCRAM}. The biological brain does have  this property to a certain degree. In fact, the brain can perform a wide variety of tasks without changing substantially the physical topology 
of its network, by simply responding to external stimuli~\cite{kohonen_book}. 

\subsection{Analog vs. digital computation}
MemComputing machines can be defined as both analog and digital (or mixed) according to the structure of the set $M$ of possible states of a single memprocessor. If that set is finite then the machine is digital. If it is a continuum or infinite 
discrete set of states then the machine operates in the analog regime. Finally, if it is some direct sum of the previous two types of sets, then the machine operates in a mixed digital-analog regime. Of course, only the digital MemComputing machines 
(DMMs) are easily scalable. The brain, instead, seems to operate mainly in the analog regime~\cite{neuro}. However, it is interesting to note that also DMMs showcase features similar to the elementary building blocks of the brain (the neurons) and their 
collective behavior. In the next Section, I will expand on this analogy.

\section{Digital MemComputing machines (DMMs)}
As already mentioned, if the set $M$ of possible states of a single memprocessor is finite, then the machine is digital. The next question is then whether such a machine can be realized in hardware. In order to answer this question, a new set of 
gates, called {\it self-organizing gates} (SOGs) have been introduced~\cite{DMM2}. These are terminal agnostic gates able to always satisfy their logical proposition irrespective of whether the incoming signals are from the traditional input or the traditional output. The key for their realization is 
the coupling of the variables of the problem---DMMs are designed to solve---with memory degrees of freedom.

\subsection{Short- and long-term memories}
To make this discussion more concrete, I report here the dynamical equations representing a DMM designed to solve for the ground state of a spin glass model Hamiltonian:
\begin{equation}\label{eqn:Ising}
	H= -\sum_{i>j} J_{ij} s_i s_j , \;\; s_i \in \{-1, 1\},
\end{equation}
where the interaction strength $J_{ij}$ (between the spin variables $s_i$ placed on a the sites of a lattice) is random, and may involve only nearest neighbor spins or any type of interaction between spins, e.g., long range. 

To design a DMM that solves such a problem, the spins are first linearized (namely they acquire a continuous value between $-1$ and $+1$), and then they are coupled to two types of memory degrees of freedom (short- and long-term memory) so that the phase space of the spin plus memory dynamics has only saddle points and equilibria representing the 
solution of the problem~\cite{Rudy}. The full set of equations is then:

\begin{equation}
		\label{eq:mem_voltages}
\begin{split}
	&\dot{s}_i = \alpha \sum_j J_{ij}s_j - 2\beta \sum_{j}x^s_{ij} s_i,\\
	&\dot{x}^s_{ij}  = \gamma C_{ij} - x^l_{ij},\; {x}^s_{ij}\in [0,1]\rightarrow\textrm{{\it short-term memory}},\\
	&\dot{x}^l_{ij}  = \delta x^s_{ij} - \zeta,\; {x}^l_{ij}\in [1,L]\rightarrow\textrm{{\it long-term memory}},
\end{split}
\end{equation}
where $C_{ij} = \frac{1}{2}(J_{ij}s_i s_j + 1)\in [0,1]$, is a clause function, and $\alpha,\beta,\gamma,\delta, \zeta$ are time-scale parameters, fixed for all system sizes, and $L$ is an arbitrary but finite upper limit for the long-term memory (see~\cite{Rudy} for the choice of these parameters and for a thorough explanation of how these equation have been derived). Note that Eqs.~(\ref{eq:mem_voltages}) can be compactly written as $\dot {\bf x}(t)=F({\bf x}(t))$, with $\bf x$ the collection 
of all variables, and $F$ the flow vector field (the right hand side of Eqs.~(\ref{eq:mem_voltages})).

The important point to make is that the short-term memory contains information on the recent history of the system dynamics, while the long-term memory contains information on the entire history. The existence and coupling of these two types of memories 
is an important ingredient to realize in practice DMMs for the solution of combinatorial optimization problems~\cite{Mybook}. 

It is interesting to note that the human brain showcases also two types of memories: long-term and short-term memory~\cite{neuro}. The short-term memory (which is believed to be mainly located in the prefrontal cortex) is assumed to be a ``working 
memory'', namely it allows us to accomplish certain tasks that may be forgotten within a relatively short time, without much detriment. Instead, the long-term memory (which is located in the hippocampus, and from there, it is supposedly distributed to the cerebral cortex) is responsible for the storing of events far in the past. It is presumably created by the reinforcement of short memories. This is similar to how the DMMs (e.g., practically realized in Eqs.~(\ref{eq:mem_voltages})) operate: the long-term memory is ``reinforced'' (via physical coupling) by the short-term one. 

\subsection{Instantons, action potentials, critical points and nodes of Ranvier}
DMMs find the solution of a given problem by traversing specific trajectories in phase space, known as instantons~\cite{topo,DMMtopo}. Instantons in DMMs are families of trajectories (a manifold) connecting a critical point (a point, $\bf x$, where the flow vector field, $F$, is zero) in phase space with another more stable critical point~\cite{Mybook}. Instantons are sudden and relatively short bursts (avalanches) of the variables around the ground potential of the system. 

Once the first instanton is initiated, it propagates the excitation to the next instanton, and so forth until the system reaches an equilibrium (if it exists). The critical points in between two successive instantons act as some sort of ``regenerative'' centers (repeaters) of the signal that propagates in the phase space, since at a critical point the system spends enough time to ``decide'' on the next instanton (trajectory) to take. 

The mechanism I just described is very much reminiscent of how action potentials (electrical polarization signals) propagate in myelinated axons. Action potentials, like instantons, quickly ``rise and fall'' around the resting potential state of the axon membrane. However, in myelinated axons they propagate in a saltatory 
fashion from a node of Ranvier to the next~\cite{neuro}. 

The nodes of Ranvier are myelin-sheath gaps along the axon where exchange of ions between the axon membrane and the environment can occur, so that the next action potential can be generated and travel along the myelinated part of the axon. This way the action potential 
can ``jump'' from one node of Ranvier to the next, allowing for a faster conduction of the signal. The critical points in the phase space of DMMs are then 
the equivalent of the nodes of Ranvier in myelinated axons.

\subsection{Long-range order}

An interesting physical property of the animal brain is the observed {\it scale-free} behavior in the firing of neurons, even in the absence of external stimuli~\cite{Munoz}. This is 
an {\it emergent} property of the collection of neurons, and it has been demonstrated in several experiments, although there is still much debate regarding its origin~\cite{Munoz}. 

For instance, experiments have shown that cortical neurons, when deposited on a grid of electrodes, fire collectively, and the size, $S$, of the neuronal avalanches (how many neurons fire together) follows a power-law distribution, $S^{-\tau}$, with $\tau$ close to $3/2$. 

This is similar to the critical Borel distribution of the size of the variable avalanches (instantons) found in DMMs solving combinatorial optimization problems~\cite{Bearden}. In fact, by means of a mean-field theory it was shown that the distribution of the size of the avalanches in DMMS is also a power-law $S^{-3/2}$, which is confirmed by numerical experiments. Together with the analogy between instantons and action potentials, this emergent property of both DMMs and the brain makes the former an 
interesting test bed to explore phenomena that could have implications on the latter. 

\subsection{Robustness against noise vs. fault tolerance}
Since DMMs employ objects of topological nature to compute (instantons) they are robust against small perturbations and noise whose strength is not enough to affect the topological structure of the phase space. However, if the architecture of the network of memprocessors is changed (by, e.g., changing even a single SOG in the circuit), the DMM would {\it not} solve the original combinatorial optimization problem it was designed for. Rather, it would attempt to solve this ``new'' problem. In other words, 
while DMMs are robust against noise and small perturbations, they are not fault tolerant. 

The brain instead seems to have both properties. In fact, it is known that neurons in the brain both die off and are generated continuously~\cite{neuro}. It is then obvious that the architecture (physical topology) of the network of neurons is not fixed 
in time. Despite these changes (provided they are not substantial), the brain continues to function as expected, namely it is both robust against noise/perturbations (the firing of single neurons in the brain still occurs) as well as topological changes: it is fault tolerant to a high degree. 

This fault tolerance could be due to the presence of time non-locality (memory) in synapses and the fact that the brain, unlike a DMM, is not attempting to solve a specific combinatorial optimization problem whose Boolean (or algebraic) expression 
is well defined. Time non-locality is a feature that allows ``re-routing'' of information ``on the fly'', despite failure of single units, as it was demonstrated in networks of resistive memories~\cite{13_self_organization}. That particular network was designed for the solution of the shortest-path problem. Memory (time non-locality) would still promote self organization of the network into the shortest possible path or paths, in the presence of defects in the network (created by eliminating some resistive memories). 

All these results seem to suggest that the more specialized the physical architecture of a network is (as for DMMs designed for specific combinatorial optimization problems), the less robust it is to topological changes. Indeed, our brains can tackle a wide variety of tasks, but are not particularly good at solving, e.g., combinatorial optimization problems. 

This fact may also be related to recent research on autism spectrum disorder (ASD). For instance, some experimental studies, employing neuro-imaging techniques, have shown structural differences in several brain regions in people with ASD compared with individuals without ASD~\cite{Autism}. These differences may be the reason children with ASD may sometimes show some mathematical skills outperforming non-autistic peers, while struggling in some other tasks. It may very well be that the brain structure of people with ASD is topologically more constrained than that of the general population. 

\section{Conclusions}
In summary, I have briefly outlined the brain-like features of MemComputing machines. Of course, these similarities do not imply that MemComputing machines are brain-like. They simply indicate that some of their dynamical properties are also observed in the operation of the biological brain. 

Of particular note is that some of these properties are {\it emerging phenonema}, namely they emerge from the collective dynamics of the units making up these machines. For instance, long-range order in DMMs---arguably the most important property for the solution of hard combinatorial optimization problems---is a feature that originates from the time non-local response of their memprocessors~\cite{Mybook}. In fact, DMMs enter this long-range ordered state ``naturally'' without tuning any parameter during dynamics. This realization may help understand the supposed critical dynamics of the brain, which is still not fully understood. Work along this direction could then be beneficial in the field of (computational) neuroscience.

\end{document}